# Self-Supervised and Supervised Deep Learning for PET Image Reconstruction

Andrew J. Reader

*School of Biomedical Engineering and Imaging Sciences, King's College London, UK*

andrew.reader@kcl.ac.uk

**Abstract.** A unified self-supervised and supervised deep learning framework for PET image reconstruction is presented, including deep-learned filtered backprojection (DL-FBP) for sinograms, deep-learned backproject then filter (DL-BPF) for backprojected images, and a more general mapping using a deep network in both the sinogram and image domains (DL-FBP-F). The framework accommodates varying amounts and types of training data, from the case of having only one single dataset to reconstruct through to the case of having numerous measured datasets, which may or may not be paired with high-quality references. For purely self-supervised mappings, no reference or ground truth data are required at all, but at minimum just the measured dataset to reconstruct from. Instead of a supplied reference, the output reconstruction from the trainable mapping is forward modelled, and the input data serve as a reference target for this forward-modelled data. The self-supervised deep learned reconstruction operators presented here all use a conventional image reconstruction objective within the loss function (e.g. maximum Poisson likelihood, maximum *a posteriori*). If it is desired for the reconstruction networks to generalise (i.e. to need either no or minimal retraining for a new measured dataset, but to be fast, ready to reuse), then these self-supervised networks show potential even when previously trained from just one single dataset. For any given new measured dataset, finetuning is however usually necessary for improved agreement with the reconstruction objective, and of course the initial training set should ideally go beyond just one dataset if a generalisable network is sought. This work presents preliminary results for the purely self-supervised single-dataset case, but the proposed networks can be i) trained uniquely for any measured dataset in hand to reconstruct, ii) pretrained on multiple datasets and then used with no retraining for new measured data, iii) pretrained and then finetuned for new measured data, iv) optionally trained with high-quality references. The overall unified framework, with its optional inclusion of supervised learning, provides a wide spectrum of reconstruction approaches by making use of whatever (if any) training data quantities and types are available for image reconstruction. Such a spectrum of reconstruction methods (ranging from purely self-supervised model-driven for a single measured dataset in hand only, through to non-model / fully-data driven) can provide a balance between a conventional reconstruction objective (e.g. data fidelity, with or without regularisation) and the potential risks/benefits of supervised regularisation (which uses training data with high-quality references).

## INTRODUCTION

Image reconstruction with deep learning has shown much promise over recent years for positron emission tomography (PET) [1]. However, there remain concerns and limitations, often related to the amount and type of training data used to train the deep networks. For supervised learning of image reconstruction (e.g. [2-5]), high quality reference data or ground truth data are required to be paired with each training measured dataset. Such reference data can be hard to obtain, or if found by simulation may not be accurately representative of real ground truth distributions. Furthermore, for medical imaging, the use of a general pool of training data is a concern, as the patient being scanned is quite potentially outside the training distribution and so there could be a risk of erroneous reconstructions, or minor deviations from the training set could result in non-trivial impacts on the final images [6].

   This work presents a unified approach with example deep reconstruction networks for PET which can help to avoid some of these problems, by combining self-supervision from unlabelled data as a core component with optional inclusion of supervised learning from a labelled database. The focus though will be on self-supervision and its benefits (while noting that self-supervision can also be easily integrated into most conventional supervised deep-learning reconstruction methods, as mentioned later in table 1). First, self-supervision of deep reconstruction operators avoids the need for high quality reference or ground truth data, through direct application of the imaging system model to the output reconstruction from the deep network, allowing the input data to serve also as the reference target in the loss function. This means, quite importantly, that real measured data can be used for training purposes, without need for simulation. Second, the self-supervised approach also leads naturally to a methodology which is suitable for any quantity of training data – from just one to many training datasets. Self-supervised training can also explicitly focus on the unique scan data in hand which are being reconstructed, which could well be an outlier relative to any training data distribution. This allows a balance between a conventional reconstruction objective (e.g. data fidelity, with or without regularisation) and a conventional deep-learned supervised regularisation task (using other training data). Seeking such a balance remains a topic of growing interest in inverse problems [7]. Figure 1 illustrates an example of this balance by contrasting conventional image reconstruction with purely deep-learned reconstruction, where it can be noted that self-supervision for a single dataset shares similarities with the model-driven approach. As a side note, self-supervised approaches can also be exploited purely as a regularisation strategy (e.g. [8], and the follow up work [9]).

Self-supervised methods have been carried out in MRI (e.g. [10, 11]) and other fields such as diffraction imaging [12]. This present work describes a unifying methodology for learning reconstruction operators (rather than purely image representations), for the case of sinograms, backprojected images (including so-called histoimages [13]) and list-mode datasets, as often encountered in PET imaging. Sinograms are of course encountered in single photon emission tomography (SPECT) as well as in transmission computed tomography (CT). The unification implicitly includes, as a special case, the deep image prior (DIP) [14] and its application to PET image reconstruction ([15], [16]) as well as CT reconstruction [17]. It is important to appreciate that previous reconstruction work with the DIP used deep networks as a means of image representation. Here, the unified description covers not only deep representations, but also considers self-supervision of deep neural networks as reconstruction operators.

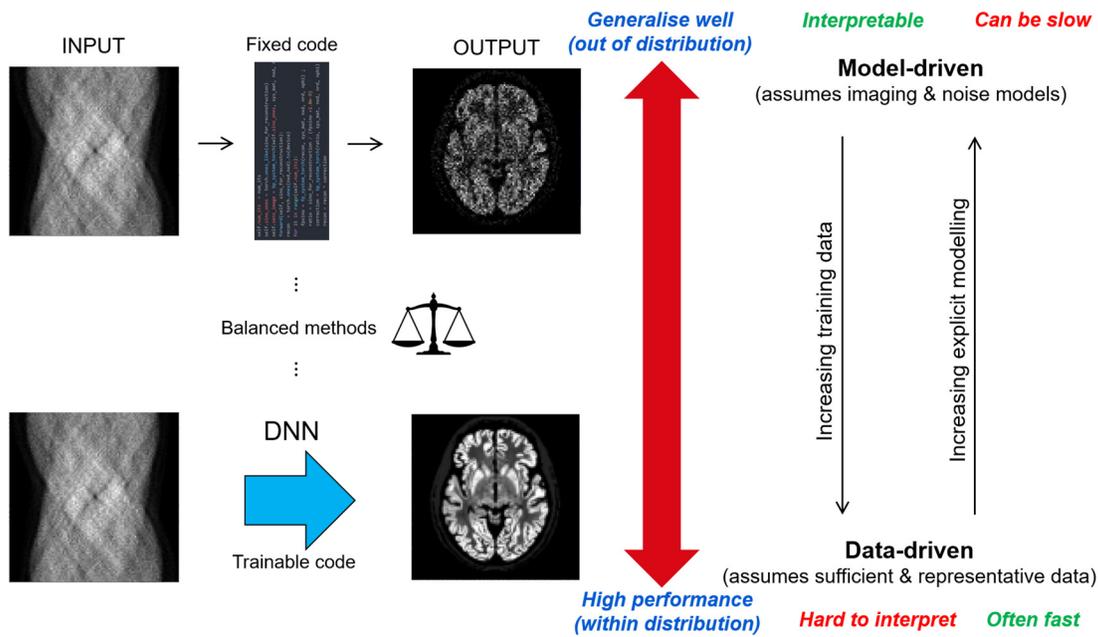

**FIGURE 1.** The top row represents the fixed mappings of conventional model-driven reconstruction methods (e.g. the first 5 methods listed later in table 1) which generalise well to unseen data, but can be slow to reconstruct (when iterative) and can be inferior in performance. The bottom row represents pure deep-learning reconstruction methods (e.g. the last 3 methods listed in table 1) using just a deep neural network (DNN) which might not always generalise well to unseen data, yet can be fast to reconstruct. The self-supervised approaches discussed in this present work share some similarities to the top-row model-driven methods when trained uniquely for the dataset in hand to reconstruct. One might seek reconstruction methods which strike a balance between the two extreme assumptions of this figure (i.e. either assuming all models are known and accurate (top row), or assuming that sufficient and representative example training data are available (bottom row)). This present work provides a unified framework that may facilitate a balance between the extremes (see example methods listed in the middle rows of table 1).

# DEEP RECONSTRUCTION WITH SELF-SUPERVISION

This section covers a unified deep-learned image reconstruction approach for PET incorporating self-supervision from sinograms or from backprojected data (e.g. backprojected list-mode data or histoimages). In the purely self-supervised case, the reconstruction network needs only the dataset in hand ($m$) to reconstruct a reliable image, and no ground truth or reference data. If, in addition, it is desired for the learned networks to generalise (i.e. to be fast and ready to use on a new dataset with no or minimal extra training), multiple augmentations of the original single dataset can be applied, but generalisation is expected to be more robust if still more measured/acquired datasets are made available, provided that sufficiently parameterised and expressive deep architectures are used within the overall network. The framework described below makes use of the imaging forward model, or system matrix, $A$, in order for the self-supervision to be possible (i.e. without need of high-quality references), and in most cases the transpose (or adjoint) of the forward model is also needed. An untested hypothesis is that only an approximate transpose is needed in such cases, whereas an accurate forward model is crucial.

## Self-Supervised and Supervised Deep Learned Filtered Backprojection (DL-FBP)

As a first example, one deep neural network (DNN) reconstruction operator is learned. The DNN operates on the measured sinogram data to deliver a processed ("filtered") sinogram, which when backprojected, delivers a reconstructed image consistent with the measured dataset $m$. This method will be referred to as self-supervised deep-learned filtered backprojection (DL-FBP) and is shown in Fig. 2. Hence a deep network operator $F$, parameterised by $\theta$, operates on the data $m$, which are then backprojected ($A^T$) to generate a reconstructed image $x$:

$$x(\theta) = a\left(\frac{A^T F(m;\theta)}{A^T \mathbf{1}}\right) \quad (1)$$

where $a(.)$ is an optional function to encourage or enforce positivity (e.g. an activation such as ReLU or PReLU, or an absolute function), and the term $A^T \mathbf{1}$ in the denominator is the sensitivity image (as encountered in methods like the expectation maximisation maximum likelihood (MLEM) algorithm [18] in PET). The sensitivity image corresponds to a backprojection of unit data (data filled with ones, $\mathbf{1}$), and simply counts the number of contributions to a given voxel or pixel when the transpose of the forward model is applied to any dataset. This helps normalise the mapping. The reconstructed image specified by equation (1) is then forward modelled to generate a corresponding model of the mean of the data, $q$:

$$q(\theta) = A x(\theta) \quad (2)$$

This model of the mean data can then be used with any chosen reconstruction objective function $D_{REC}$, whether regularised or unregularised, such as least squares (LS), maximum likelihood (ML) or maximum *a posteriori* (MAP). For PET image reconstruction a common choice is the Poisson log-likelihood for data consistency (either with, $\lambda > 0$, or without, $\lambda = 0$, a regularising prior, $R$),

$$D_{REC}(q(\theta); m) = -\sum_{i=1}^{I}\bigl(m_i \ln q_i(\theta) - q_i(\theta)\bigr) + \lambda R(\theta) \quad (3)$$

where the negative allows expression of this objective function as a loss function to be minimised by optimising $\theta$. Optimisers used in deep learning, such as Adam [19], are effective in optimising the parameters $\theta$ of the sinogram-domain DNN $F$, so as to find a minimum of equation (3). For the case of using the negative Poisson log-likelihood for the reconstruction objective function, it is important to constrain the image generating equation (1) to be positive, which can be done, for example, by using an absolute (so $a(.) = |.|$ in equation (1)), and the learning of the parameters for network $F$ will compensate as required for the use of $a(.)$.

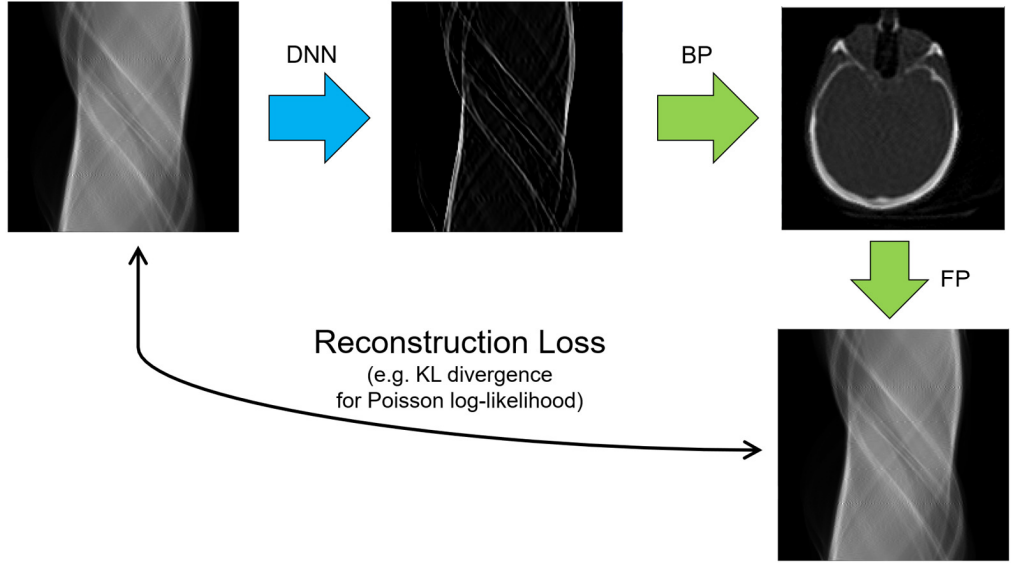

**FIGURE 2.** Simplified schematic of the processing pipeline for self-supervised deep-learned filtered backprojection (DL-FBP), where the reconstruction loss could be least squares or, for example, the Kullback-Leibler (KL) divergence (with optional regularisation). DNN denotes a deep neural network (the blue colour indicates it is trainable), BP denotes backprojection and FP denotes forward projection (green colour indicates these are fixed mappings). The filtered sinogram after the DNN has the appearance of having been processed with a high-pass filter, as indeed would be expected given that standard FBP would use a high-pass ramp filter of the sinogram prior to backprojection. A more general framework is shown in Fig. 3.

The following observations can be made for equations (1) to (3):

a) If the forward imaging model $A$ is just a discrete x-ray or Radon transform, and if the loss function is simply the mean square error (MSE), then a self-supervised deep learned version of filtered backprojection (FBP) is obtained. (Python / PyTorch code for this very implementation has been published in an online video by the author [20], and was first included in a May 2022 conference invited talk [21].)

b) Equation (1) models the reconstructed image as a combination of the rows of the system matrix $A$, followed optionally just by a simple function $a(.)$. Hence the DNN is estimating coefficients for each of the rows of $A$ in order to combine them as spatial basis functions to synthesise a reconstructed image. For the Radon transform these spatial basis functions correspond to images of lines of response (LORs) associated with each sinogram bin. For noisy data, this can have a potentially helpful regularisation effect, depending on the rank of matrix $A$, compared to the case of estimating coefficients of pixel or voxel spatial basis functions directly.

c) Any chosen statistical model appropriate to the data, such as the Poisson noise model in (3), can be used. If the loss function is the negative Poisson log likelihood only (no prior $R$), we obtain what could be referred to as "*maximum Poisson likelihood FBP*". This would, after many training epochs, deliver results comparable to methods such as MLEM when run for many iterations, as demonstrated in the results section below.

d) Any chosen forward imaging operator $A$ (including accurate imaging models if available) can be used, yet the transpose of the system model, $A^T$, could potentially be imprecise, or not even used (as will be explicitly considered below). This is because the DNN will learn to compensate for approximations in the backprojection. In the extreme case a DNN can bypass the need for $A^T$ altogether, but at great cost to generalisation. Likewise, if data are missing, dedicated self-supervision ensures that the generated image $x$, when mapped through the forward model $A$, agrees with the data. Hence the primary need is for an accurate forward model $A$.

e) There is no need for any ground truth reference or high-quality reference data, as only the unique data in hand to reconstruct are used in equation (3). For the single dataset case, a unique and appropriate reconstruction operator for the single dataset will be found by minimising (3). As will be seen later, augmentations of the single dataset can drive the learning towards a more general reconstruction operator.

To make a more general unified approach, the overall loss function can be augmented to include i) the self-supervised component $D_{REC}$ for the reconstruction objective for the data in hand, ii) a self-supervised component $D_{NOREF}$ for any other available data which have no associated reference, as well as finally, iii) a supervised component $D_{REF}$ if appropriate reference training data are available (e.g. high count reference reconstructions, or knowledge of ground truth images, each paired with a standard quality measured dataset). In this case, the total loss function for the optimisation of $\boldsymbol{\theta}$ becomes

$$L_{TOTAL}(\boldsymbol{\theta}) = \alpha D_{REC}(\boldsymbol{q}(\boldsymbol{\theta}); \boldsymbol{m}) + \beta D_{NOREF}(\{\boldsymbol{q}_{NR}(\boldsymbol{\theta}); \boldsymbol{m}_{NR}\}) + \gamma D_{REF}(\{\boldsymbol{x}_R, \boldsymbol{m}_R\}) \tag{4}$$

where $\alpha$ allows a weighting for the scan-data unique self-supervised reconstruction loss, $\beta$ provides weighting for any available unlabelled training set (just data, with no references $\{\boldsymbol{m}_{NR}\}$, which can also include augmentations of the data in-hand, $\boldsymbol{m}$), and finally $\gamma$ gives a weight for the supervised loss based on a training set of one or more pairs of high quality reference datasets (high quality reference image $\boldsymbol{x}_R$, each paired with a dataset $\boldsymbol{m}_R$).

Equation (4) allows control over the balance between a conventional reconstruction objective (according to the size of $\alpha$) and a supervised objective using high-quality references (according to the size of $\gamma$). Many other variations are possible – e.g. just training with $\alpha=1$ ($\beta=0$, $\gamma=0$) for the data in hand, or pretraining with $\beta=1$, $\gamma=1$ ($\alpha=0$), and then using with no further training or else finetuning with $\alpha=1$. It is worth noting that if the first two terms of (4) are disregarded (i.e. $\alpha=0$, $\beta=0$), and if only one reference pair of data are available, then in essence the very early 1991 work of Floyd [22] is obtained, which relied on supervised learning from reference data to learn the parameters for a convolution kernel in the sinogram domain. Similarly, a variety of conventional supervised learning cases are represented by the case of $\alpha=0$, $\beta=0$, such as the FBP-net method [23], which uses just an MSE loss for supervised learning only. Figure 3 summarises the proposed general approach of equations (1) to (4) but also importantly extends it further with an additional image-space DNN in accordance with the following section (and equation (5)).

## Self-Supervised and Supervised DL-FBP Followed by Image-Space Filtering (DL-FBP-F)

The method of equations (1) to (4) can be extended in an important way to include a second DNN, this time applied in image space. The method will be referred to as DL-FBP-F, to reflect the additional processing "filter" in the image domain. Equation (1) is replaced by:

$$\boldsymbol{x}(\boldsymbol{\theta}_1, \boldsymbol{\theta}_2) = \boldsymbol{F}_2\left(a\left(\frac{\boldsymbol{A}^T \boldsymbol{F}_1(\boldsymbol{m}; \boldsymbol{\theta}_1)}{\boldsymbol{A}^T \boldsymbol{1}}\right); \boldsymbol{\theta}_2\right) \tag{5}$$

where now there is an image-space DNN $\boldsymbol{F}_2$, operating on the deep-filtered backprojection (which has first been processed by an appropriate function $a(.)$, such as in this case a parametric ReLU (PReLU)). Equation (5) enables a network to generate unique coefficients for each and every pixel or voxel spatial basis function, and so has greater scope for minimising the reconstruction objective compared to equation (1). This leads to more able agreement with the chosen reconstruction objective function, and is an important advantage of DL-FBP-F, compared to DL-FBP. However, greater scope for data fidelity can also mean greater levels of image noise if the data are noisy and if a non-regularised objective function is used. As before, this reconstructed image is forward modelled to generate a corresponding model of the mean data, $\boldsymbol{q}$:

$$\boldsymbol{q}(\boldsymbol{\theta}_1, \boldsymbol{\theta}_2) = \boldsymbol{A}\boldsymbol{x}(\boldsymbol{\theta}_1, \boldsymbol{\theta}_2) \tag{6}$$

Note again that there needs to be an absolute or similar positivity constraint applied to $\boldsymbol{Ax}$, if objective functions such as the Poisson log likelihood are to be used within $D_{REC}$. Hence two deep mappings, with parameters $\boldsymbol{\theta}_1, \boldsymbol{\theta}_2$, need to be optimised. Other variations include operating on backprojected data, to give DL-BPF (deep-learned backproject then filter):

$$\boldsymbol{x}(\boldsymbol{\theta}) = a\left(\frac{\boldsymbol{F}(\boldsymbol{A}^T\boldsymbol{m}; \boldsymbol{\theta})}{\boldsymbol{A}^T\boldsymbol{A}\boldsymbol{1}}\right) \tag{7}$$

where the reconstruction loss can be optionally applied in the backprojected image space (comparing between $\boldsymbol{A}^T\boldsymbol{m}$ and $\boldsymbol{A}^T\boldsymbol{A}\boldsymbol{x}(\boldsymbol{\theta})$) or indeed still in the sinogram domain as before. Using a loss based on the backprojected image could be easier for histoimages for example.

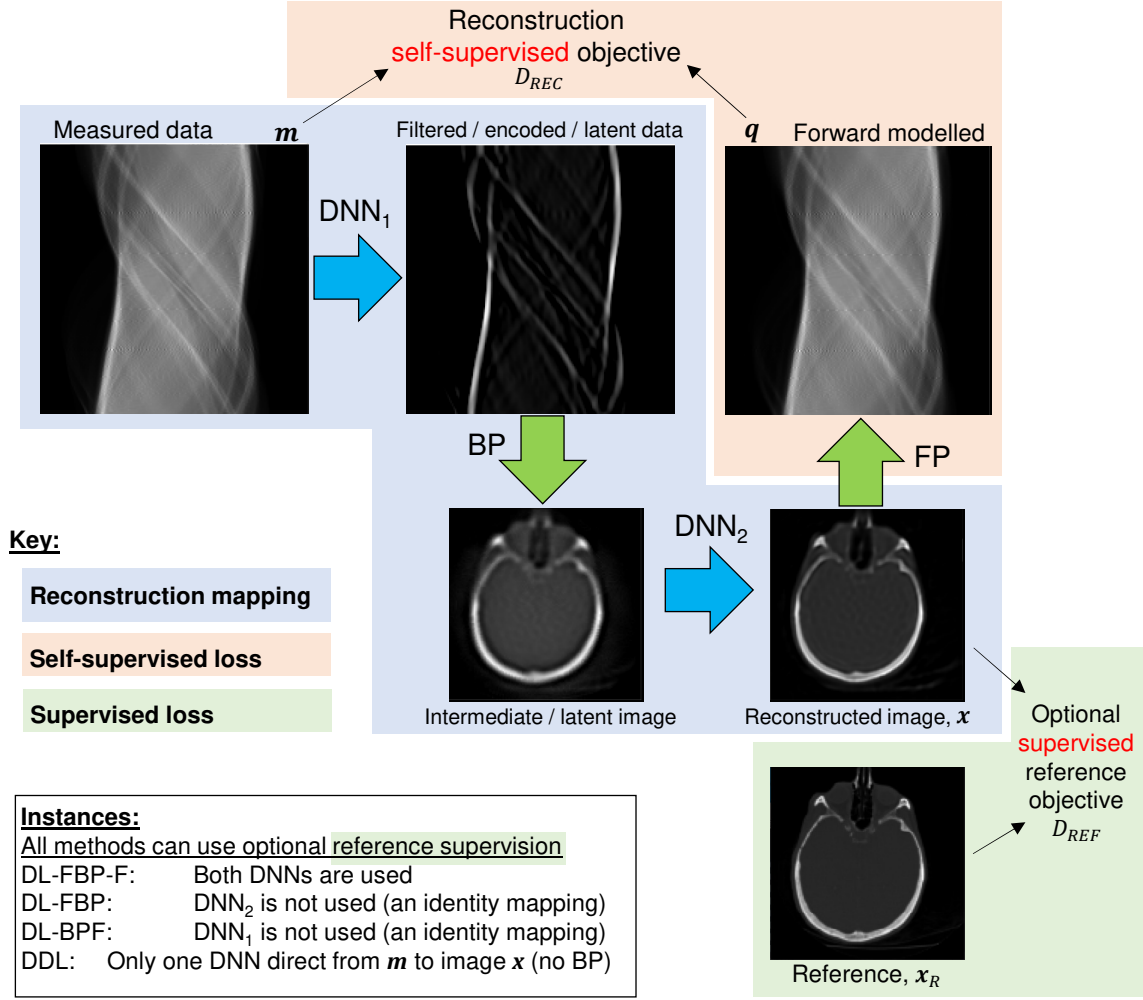

**FIGURE 3.** Unified self-supervised and supervised deep-learning reconstruction framework, including deep-learned filtered backprojection (DL-FBP) and the case with an additional image-space deep network (DNN$_2$), called DL-FBP-F. DNN$_1$ can be any deep-learned network sufficient for processing the sinogram, such that if backprojected it forms an image, which ultimately when forward projected agrees with the original data. The framework is more general than depicted, as the two main networks (DNN$_1$ and DNN$_2$) can be a more general encoder-decoder, to consider latent spaces for differing feature maps and sizes. Table 1 covers a more general description.

A further, simple approach is a direct deep learning (DDL) method:

$$x(\theta) = a\big(F(m; \theta)\big) \tag{8}$$

which, while avoiding use of the system model in the generated reconstruction (reducing assumptions about the transpose of the system model), poses a greater challenge if the method is to generalise beyond just one training set. In fact, methods like reconstruction using DIP [16] and even DeepPET [2] can be regarded as differing instances of a model such as (8). In the case of the purely self-supervised DIP, noise or an image is supplied instead of the measured data $m$ (DNNs can be so expressive that they can compensate for nearly any input), and for DeepPET the loss function only consists of data paired with references (labels), hence a purely supervised loss with $\alpha = 0, \beta = 0$ and $\gamma = 1$ in the total loss function of equation (4).

A finishing note for this section is to mention the link to model-based deep learning [24] as well as "unrolled" reconstruction methods, such as the learned primal dual (LPD) architecture [25], [26], the iterative neural network [27] and FBSEM-Net [5] as examples. These methods can be regarded as model-based deep learning in the sense that they use the system model within their mappings. Such mappings could also be extended in the self-supervised way presented here, simply by applying the forward model to the output reconstruction, allowing the forward model of the reconstructed image to be compared to the input data in the loss / objective function. Or indeed they can be trained in a supervised way only (as in their originally presented and intended form), or indeed via a weighted combination of both self-supervised and supervised learning (using the loss function of equation (4)).

Table 1 (in tandem with Fig. 3) shows how the proposed framework (with its four example networks: DL-FBP, DL-FBP-F, DL-BPF and DDL) is situated compared to many other image reconstruction methods.

**TABLE 1.** PET reconstruction methods including those proposed here (DL-FBP, DL-FBP-F, DL-BPF) showing how all DL methods can be extended via inclusion either of self-supervised or supervised components. Colour coding relates to Fig. 3, where blue indicates the reconstruction mapping, orange represents use of self-supervised components and green indicates use of supervised components (including reference images, $x_R$). All the trainable mappings (i.e. with $\theta$, in red) can be extended to accommodate both types of supervision. BP denotes backprojection ($A^T$), FP denotes forward projection ($A$).

| Recon method name | Input data | Pre-process | Encoder TRANSFORM Input to latent → | Latent space TRANSFORM Latent to latent → | Decoder TRANSFORM Latent to output → | Post-process | Output recon | Components in mapping from $m$ to recon $x$ $\theta$ trainable | Components in main loss or objective (for method as defined) | Optional / proposed extension to loss |
|---|---|---|---|---|---|---|---|---|---|---|
| Pseudo inverse of $A$ | $m$ | | Matrix of left singular vectors of $A$ | Inverse of matrix of singular values | Matrix of right singular vectors of $A$ | | $x$ | $A$ | $A, m$ | |
| FBP | $m$ | | FT | Ramp filter | Inv. FT | BP | $x$ | $A$ | $A, m$ | |
| BPF | $m$ | BP | FT | Reciprocal of FT of PSF | Inv. FT | | $x$ | $A$ | $A, m$ | |
| CBP | $m$ | | | Convolution kernel | | BP | $x$ | $A$ | $A, m$ | |
| MLEM [18] | $m, x^{(k)}$ | | FP | Expectation of complete latent data z | BP (for max.) | | $x^{(k+1)}$ | $A, m$ | $A, m$ | |
| DIP [16] | $z$ | | | (input is fixed latent $z$) | DNN | | $x$ | $\theta$ | $A, m$ | |
| DL-FBP | $m$ | | | DNN | | BP | $x$ | $A, \theta$ | $A, m$ | $x_R$ |
| DL-FBP-F | $m$ | | DNN | BP | DNN | | $x$ | $A, \theta$ | $A, m$ | $x_R$ |
| DDL | $m$ | | | DNN | | | $x$ | $\theta$ | $A, m$ | $x_R$ |
| DL-BPF | $m$ | BP | | DNN | | | $x$ | $A, \theta$ | $A, m$ | $x_R$ |
| Floyd [22] | $m$ | | | Learned convolution kernel | | BP | $x$ | $A, \theta$ | $x_R$ | $A, m$ |
| FBSEM-Net [5] | $m$ | | CNN in parallel with EM update, fused by maximisation step… repeated… | | | | $x$ | $A, m, \theta$ | $x_R$ | $A, m$ |
| LPD [25,26] | $m$ | | DNN    BP    DNN    FP    DNN    …    repeated… | | | | $x$ | $A, m, \theta$ | $x_R$ | $A, m$ |
| DeepPET [2] DPIR-Net [4] AUTOMAP [3] | $m$ | | DNN encoder (space to features) | CNN | DNN decoder (features to space) | | $x$ | $\theta$ | $x_R$ | $A, m$ |

## Generalisation, Self-Augmentation and Finetuning

If more measured datasets are available, the reconstruction networks such as DL-FBP and DL-FBP-F can be assisted towards general applicability with either no, or very limited, retraining needed for any new test dataset. For any number $n = 1 \ldots N$ of datasets (with no known ground truths nor reference images), an increasingly general parameterisation $\theta$ of the overall reconstruction network can be found by including as much training data as is available in the loss function, as previously included in equation (4):

$$D_{NOREF}(\{q^n_{NR}(\theta); m^n_{NR}: \ n = 1 \ldots N\}) = -\sum_{n=1}^{N}\left(\sum_{i=1}^{I} m^n_i \ln q^n_i(\theta) - q^n_i(\theta)\right) \quad (9)$$

Using equation (9) within the overall loss function (equation (4) with $\beta > 0$) seeks a parameterisation $\theta$ of the overall deep reconstruction network with its operator(s), which best matches all the data in the training set, importantly working even for just one dataset ($n = 1$), under the condition that the one or more DNNs have adequately expressive inductive priors. For example, while for just one single dataset an overparameterised simple, shift-equivariant convolutional neural network (CNN) may suffice through overfitting, for a more generally applicable network, trained from many datasets, shift-variant mappings will in general be necessary (e.g. by use of downsampling and upsampling within a CNN, to eliminate shift-equivariance). In this work, a simple self-augmentation method will be shown to be capable of achieving a useful level of generalisation from just one single dataset. In this case, the multiple training data sets in (9) are merely augmentations of the single input dataset $\boldsymbol{m}$, obtained by random rescaling, generating new noise realisations (e.g. by taking each value $m_i$ of $\boldsymbol{m}$ as a parameter of a Poisson distribution and taking a sample), and randomly removing some projection bins of data by setting them to zero (augmentation by missing data). More details of the self-augmentation are covered below in the methods section. When presented with new data to reconstruct, a finetuning approach can optionally be used: simply initialise with the last best known trained $\boldsymbol{\theta}$, and then retrain/finetune the parameters in accordance with the new single dataset to reconstruct.

## METHODS

The self-supervised learning aspect of the unified methodology was first assessed for the simple case of one single dataset for reconstruction, with no additional data. The second stage of assessment explored generalisation of the deep learned networks, again using only a single dataset, but now through exploiting self-augmentation strategies to assist in generalisation.

### Single Dataset Self-Supervised Reconstruction

Some examples of deep reconstruction networks using the proposed methodology of Fig. 3 were assessed with a simple system model $\boldsymbol{A}$ implemented in Python with the PyTorch deep learning framework. A nearest neighbour discrete 2D x-ray transform was used for the system model $\boldsymbol{A}$, applied for fast tests with a 96×96 image / test phantom. The 2D test phantom was a resized 96×96 slice of the head CT data included in the scikit image processing library (using a subset of the datasets in the "University of North Carolina Volume Rendering Test Data Set" archive). The system matrix mapped the 96×96 image to a 96×96 sinogram (96 radial bins, with 96 azimuthal viewing angles). The true object and an example noisy sinogram (Poisson noise was used in all tests), along with an example MLEM reconstruction is shown in Fig. 4. In addition, Fig. 4 shows a test dataset which was not used at any point for network training, but which allowed (where appropriate) network generalisation to be at least partly assessed by checking test-time performance on this data. Just one architecture was considered for the DNNs, simply a non-linear shift-equivariant mapping via a CNN (variable layers, with varying channel numbers, 9×9 kernels, and with PReLU activations between the convolutional layers). In the below, the number of layers reported for a given CNN correspond to the number of innermost layers (excluding the first layer which maps from one 2D sinogram or image array to a number of feature maps corresponding to the number of channels, and excluding the last layer which maps from multiple feature maps back to one single output 2D array). The Adam optimiser was used throughout, with a learning rate of $5 \times 10^{-6}$ in all of the tests. The normalised root mean square error (RMSE) and the Poisson log likelihood (PLL) were used to assess image quality and the reconstruction objective function respectively. The PLL was used as the reconstruction objective function throughout, enabling direct comparisons with MLEM which is known to maximise the Poisson log-likelihood when sufficiently iterated.

### Single Dataset Self-Augmentation for Self-Supervised Learning of Image Reconstruction

The second stage of tests concerned generating unique variations of the single measured dataset, and training the reconstruction networks in an autoencoding self-supervised fashion. The self-augmentation strategy involved random selection of one of three data modifications at each training epoch. Each strategy had equal probability of being selected at any given training epoch. Strategy one involved randomly rescaling the input sinogram data by a factor (0,10], and then generating a Poisson realisation from the sinogram data (even if the starting data are themselves noisy). Strategy two involved randomly rescaling the sinogram data and then randomly removing some of the sinogram bin values. Strategy three involved doing both of the former strategies simultaneously (removing bins, random rescaling, and introducing new Poisson noise). Hence infinitely many possible variations of the input data were able to be generated, providing a unique input (but same target) sinogram for any given training epoch of the reconstruction network. The assessments were conducted with a 96×96 sampled version of the phantom previously described. In addition, as generalisation of the learned reconstruction operators was of interest, completely unseen (by the training) test data were also reconstructed. The test data were formed from the BigBrain PET-MR phantom data [28]. The self-supervised methodology was assessed for the DL-FBP, DL-FBP-F and DDL methods.

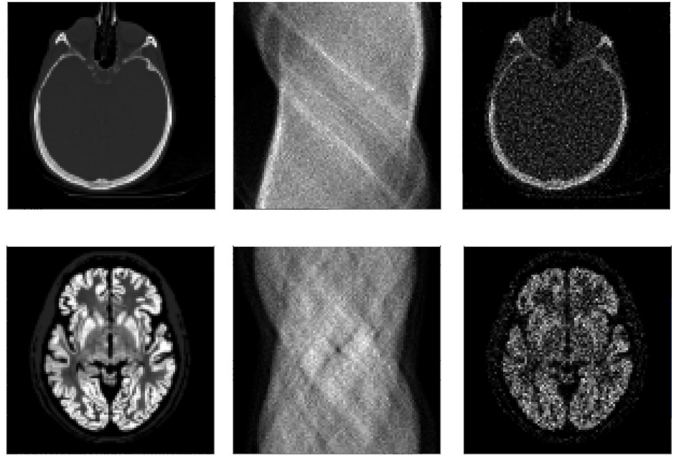

**FIGURE 4.** Top row shows the CT 96×96 phantom, and its noisy sinogram (96 bins with 96 view angles) used for training, and an example MLEM reconstruction at >200 iterations. The bottom row shows the PET 96×96 phantom used as challenging different test data case, its noisy sinogram and an example MLEM reconstruction of the test noisy sinogram data.

## RESULTS

In the first part of the results, exploratory performance of DL-FBP-F, DL-FBP and DDL for just one training / reconstruction dataset is considered. In the second part of the results, generalisation by self-augmentation is considered.

### Basic Comparison of Methods: One Dataset

Figure 5 compares the DL-FBP, DL-FBP-F and DDL methods for approximately the same number of trainable parameters (~12 M, for an overparameterised CNN), revealing that DL-FBP is the slowest of all the methods to converge towards a maximum Poisson log-likelihood estimate (represented by MLEM), with DL-FBP-F being faster, and DDL being the fastest of all (in terms of number of epochs). To achieve a comparable number of trainable parameters for each of the three methods, the two DNNs used for DL-FBP-F each had just half as many layers as the single DNNs used for DL-FBP and DDL. It is clear that all three methods deliver very comparable results to MLEM, with DL-FBP-F needing fewer epochs than DL-FBP. Figure 6 shows the result of testing these three trained networks for unseen new test data (previously shown in Fig. 4). As can be expected, testing on new unseen data does not perform well, as each of the trained networks has taken on *representation* learning (with the simple overparameterised CNNs) rather than general reconstruction *operator* learning. As can be seen, the DDL method has the poorest level of generalisation, delivering a test-time result which still resembles a sinogram rather than an image. The network has largely learned a unique representation of the unique training data rather than a general operator for mapping from a sinogram to an image, and hence it fails for new test data. This is kind of failure is what might be expected of the deep image prior when applied to new test data with no retraining. In contrast, the DL-FBP-F method shows promise for generalisation to unseen test data, clearly benefitting from the incorporation of the backprojection operator (i.e. including the imaging physics model) within the overall network.

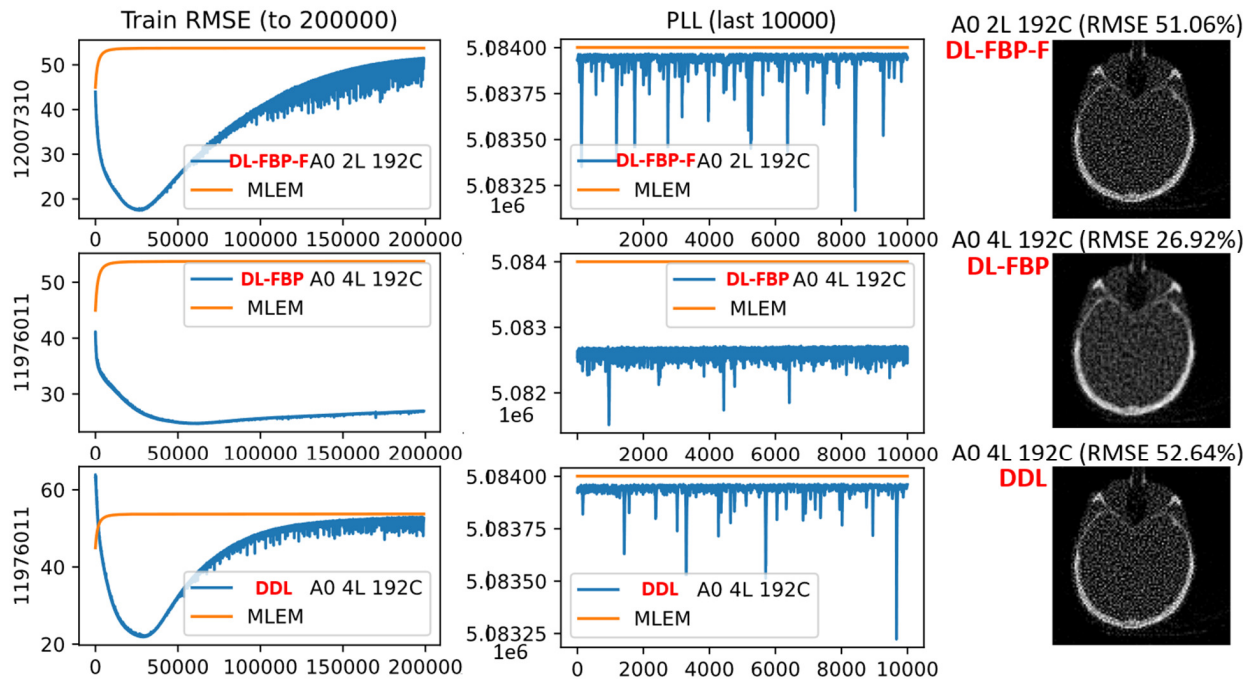

**FIGURE 5.** (a) Example results for DL-FBP-F, DL-FBP and DDL. The RMSE and the Poisson log likelihood (PLL) as a function of training epoch for one single sinogram of training data (no augmentation: "A0") is shown for up to 200k epochs (only the last 10k epochs are shown for PLL). The end point reconstructions for the more quickly converging DL-FBP-F and DDL are both close to the MLEM reconstruction (shown in Fig. 4). After 200k iterations, the RMSE for MLEM was 53.74%, visually indistinguishable from the DL-FBP-F and DDL reconstructions. The deep networks used here where CNNs with 9×9 kernels. The single CNN used for DL-FBP and DDL had 4 innermost layers ("4L") with 192 channels ("192C") per layer. As DL-FBP-F had 2 CNNs, each CNN had 2 innermost layers ("2L") to give comparable trainable parameters to the other methods. The total number of trainable parameters is reported in the leftmost column as vertical text (~12 million parameters for each method).

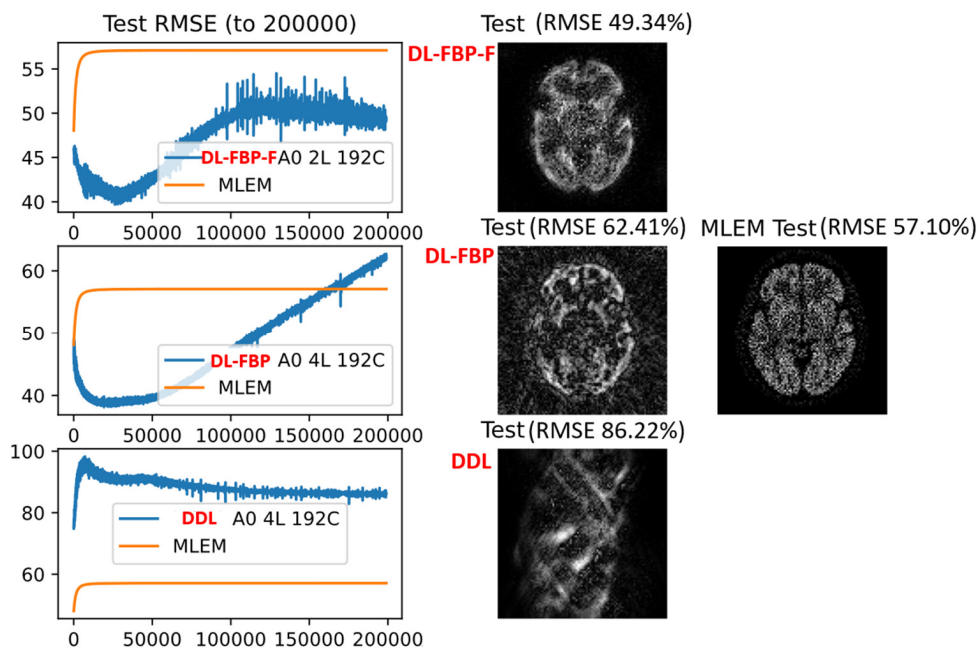

**FIGURE 6.** Anecdotal example test-data results for the networks trained in Fig. 5, shown for when test data is separately and independently passed through the current state of each network at each epoch (during training for the dataset shown in Fig. 5). The result after 200k epochs shows that DDL has the worst ability to generalise, whereas DL-FBP-F perhaps has potential to generalise. The MLEM result for the test data is shown on the right as a reference indicator for the maximum PLL estimate.

## Comparison of Methods: One Dataset with Training Augmentation

Figure 7 compares the three methods (DL-FBP-F, DL-FBP and DDL) for the case of single dataset augmentation. Using augmentation of the single training dataset leads to a notable improvement in test-time performance compared to Fig. 6. Even though only one dataset was used for training, the multiple self-augmentations encourage the reconstruction networks towards learning of a reconstruction *operator*, rather than merely a representation of the unique data in hand at training time. The test-time results are comparable to underconverged MLEM reconstructions, but clearly show favorable characteristics (such as a lower RMSE) compared to the highly iterated MLEM maximum PLL estimate. Figure 7 also shows the results of continued single-dataset non-augmented training of the networks found after the augmented training. Hence whilst the networks can potentially be used in isolation with no further training (based on the example test-time data reconstruction performance), they can of course also be finetuned to any unique data to reconstruction from (as indicated by "finetuning" in Fig. 7). In the finetuning case, the networks can clearly approach the maximum PLL estimate, although in practice this is a highly noisy estimate and rarely sought, hence the suggestion that the networks trained on just one self-augmented dataset are worth researching further to investigate potential use with no further training nor finetuning. Note that the DDL method, even with self-augmented training, is unable to deliver a generalised operator, delivering still a poor test-time reconstruction. Hence the inclusion of the imaging system model within the reconstruction mapping (through the backprojection step in DL-FBP and DL-FBP-F) is clearly crucial for delivering reconstruction networks with greater capacity for generalisation to unseen data.

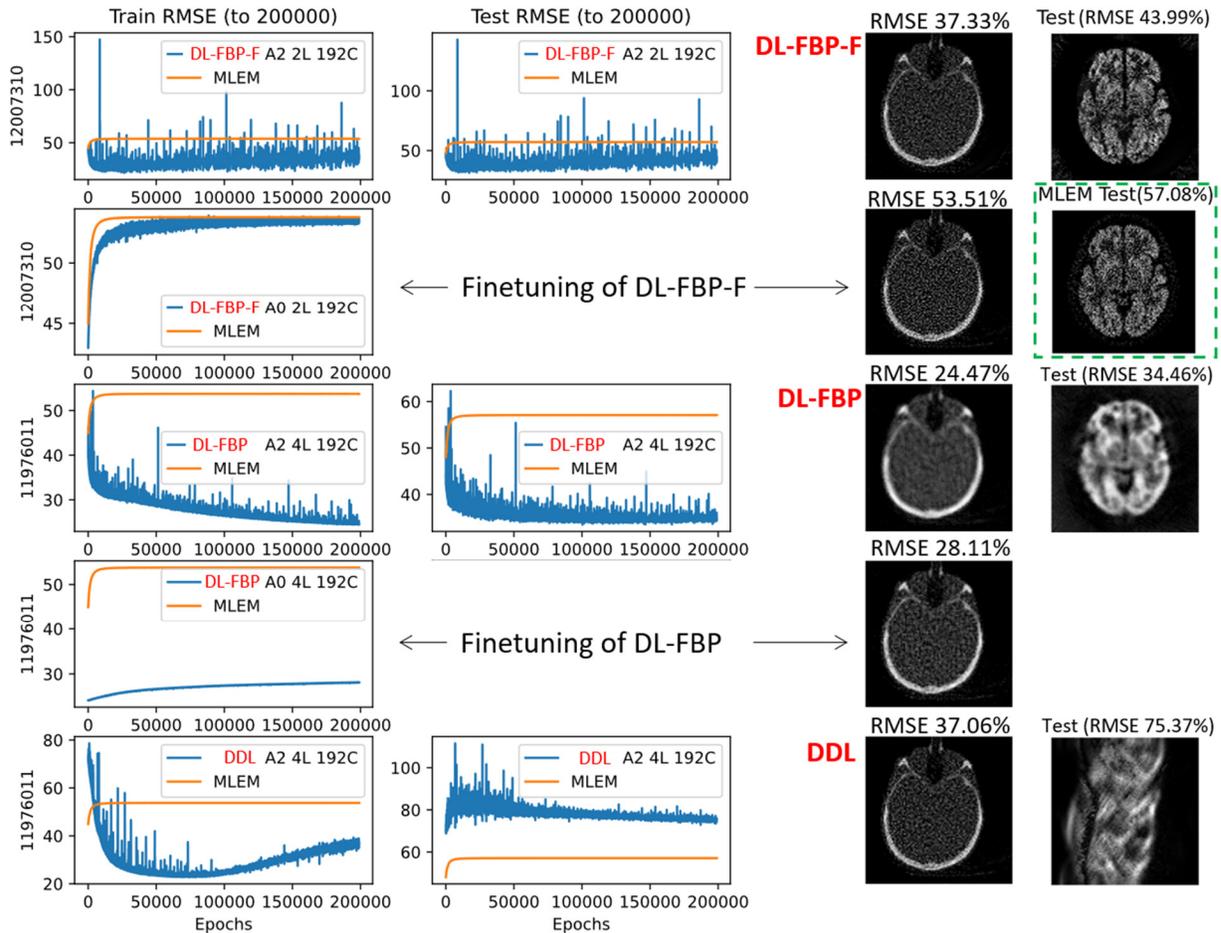

**FIGURE 7.** Comparison of the methods with training augmentation ("A2"). Test data results are improved, except for DDL which does not use the imaging system model within its mapping. The reference for the test reconstructions is the MLEM reconstruction, shown in the dashed box on the rightmost column. The total number of trainable parameters is reported in the leftmost column as vertical text (~12 million parameters for each method).

# Comparison of Impact of Architecture for DL-FBP-F

Figure 8 compares the impact of the number of channels per convolutional layer and the number of innermost layers used in each of the two CNNs used for the DL-FBP-F network. Increasing the number of channels leads to fewer epochs being needed to reach convergence (although this comes at greater computational burden per epoch). In a similar fashion, increasing the number of innermost layers in each of the CNNs used in the network also leads to fewer epochs being required to reach convergence – although this comes not only at greater computational burden per epoch, but also at potentially increased instability in the optimisation, as anecdotally evidenced in the figure (the training RMSE is more erratic than for other plots).

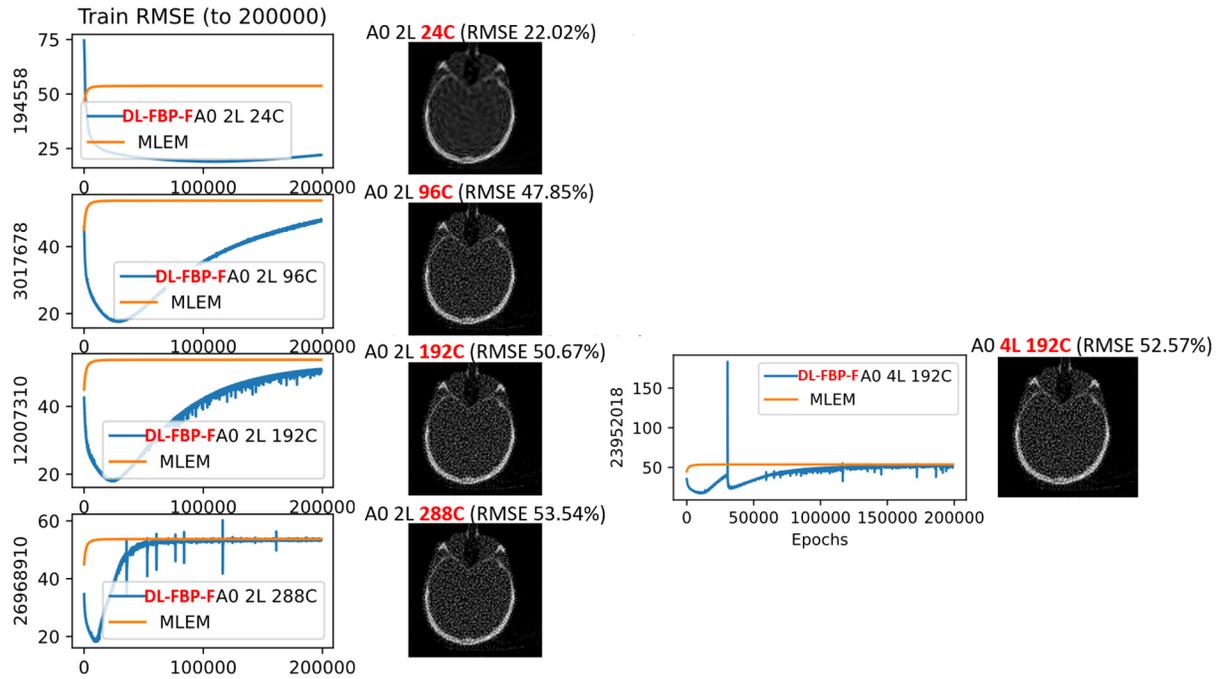

**FIGURE 8.** Columns 1 and 2 compare the number of channels (from 24 up to 288, as indicated in red above each image) and row 3 compares the number of innermost layers (from 2 to 4) for self-supervised training of the DL-FBP-F network. The total number of trainable parameters in each case is reported on the vertical axis of each plot.

# CONCLUSION

A unification of self-supervised deep learned PET image reconstruction with conventional supervised methodology is described, with particular emphasis on the general framework shown in Fig. 3. This allows any amount of training data to be used (even just the one dataset needed for reconstruction), and it does not need any high-quality reference or ground truth data. Whilst the preliminary results shown here suggest efficacy with self-augmentation of just a single dataset, the expectation is that as more training data are considered, the deep operators and networks will become increasingly general through pretraining, being applicable with only limited finetuning or potentially no retraining at all for a new test unseen dataset. Such pretrained networks would deliver fast reconstruction operators and networks for any reconstruction objective function (which usually needs slower iterative optimisation). Of course, for such generalisation to occur, careful selection of the DNN architectures used with the networks will be needed (e.g. if CNNs are used, downsampling or upsampling would likely be necessary to avoid overly simple shift-equivariant mappings).

The proposed reconstruction networks can either be i) trained uniquely for the measured data in hand, ii) pretrained on multiple datasets and then used with no new training for a new dataset, or iii) pretrained and then finetuned to the unique data to reconstruct. The method is general for any imaging system and any reconstruction objective function, and furthermore each approach can also accommodate supervised training data (i.e. use of high-quality paired references). This present work though has outlined example reconstruction operators and networks, demonstrating reconstruction with just one training dataset (i.e. the data to reconstruct from) for 2D simulated sinogram data and using self-supervision only. Further work will be needed to assess the impact of more training data, deep architecture choices, generalisation capabilities (with and without finetuning), inclusion of regularisation in the reconstruction objective and regularisation via use of a supervised component in the total loss (making use of high-quality reference data). Using the backprojection operator in DL-FBP and DL-FBP-F clearly helps the generalisation of the overall inductive prior of each reconstruction network. The direct method (DDL) would demand an architecture which is more complex, as it effectively also needs to learn a sinogram to image domain mapping (i.e. just like DeepPET).

## ACKNOWLEDGMENTS

This work was supported in part by the Wellcome/EPSRC Centre for Medical Engineering [WT 203148/Z/16/Z], and in part by EPSRC grant number [EP/S032789/1]. For the purposes of open access, the author has applied a Creative Commons Attribution (CC BY) licence to any Accepted Author Manuscript version arising, in accordance with King's College London's Rights Retention policy. The data that supports the findings of this study are available within the article with simulation data from https://scikit-image.org/docs/stable/api/skimage.data.html, https://pet-mr.github.io/brain_phantom/ and with a coding example in this video https://youtu.be/WgG-TF8XqIA.